\documentclass[journal=jpclcd,manuscript=letter]{achemso}

\usepackage{chemformula} 
\usepackage[T1]{fontenc} 
\usepackage{float}
\usepackage{dcolumn}
\usepackage{graphicx}
\usepackage{siunitx}
\usepackage{amsmath}
\usepackage{amssymb}
\usepackage{braket}
\usepackage{float}
\usepackage{graphicx}
\usepackage{siunitx}
\usepackage{amsfonts}
\usepackage{color}
\usepackage{soul}
\usepackage{braket}
\usepackage{verbatim}

\newcommand{\beq}{\begin{equation}}
\newcommand{\eeq}{\end{equation}}

\author{Giovanni Batignani}
\affiliation[Rome]
{Dipartimento di Fisica,~Universit\'a~di~Roma~``La Sapienza",  ~Roma, ~I-00185, ~Italy}
\alsoaffiliation[IIT]
{~Istituto~Italiano~di~Tecnologia, Center for Life Nano Science @Sapienza, Roma, ~I-00161,  ~Italy}
\email{giovanni.batignani@uniroma1.it}
\author{Carlotta Sansone}
\affiliation[Rome]
{Dipartimento di Fisica,~Universit\'a~di~Roma~``La Sapienza",  ~Roma, ~I-00185, ~Italy}
\author{Carino Ferrante}
\affiliation[Rome]
{Dipartimento di Fisica,~Universit\'a~di~Roma~``La Sapienza",  ~Roma, ~I-00185, ~Italy}
\alsoaffiliation[IIT]
{~Istituto~Italiano~di~Tecnologia, Center for Life Nano Science @Sapienza, Roma, ~I-00161,  ~Italy}
\author{Giuseppe Fumero}
\affiliation[Rome]
{Dipartimento di Fisica,~Universit\'a~di~Roma~``La Sapienza",  ~Roma, ~I-00185, ~Italy}
\author{Shaul Mukamel}
\affiliation[Rome]
{Department of Chemistry, University of California, Irvine, 92623, California, USA}
\author{Tullio Scopigno}
\affiliation[Rome]
{Dipartimento di Fisica,~Universit\'a~di~Roma~``La Sapienza",  ~Roma, ~I-00185, ~Italy}
\alsoaffiliation[IIT]
{~Istituto~Italiano~di~Tecnologia, Center for Life Nano Science @Sapienza, Roma, ~I-00161,  ~Italy}
\email{tullio.scopigno@uniroma1.it}

\title{Excited-state energy surfaces in molecules revealed by  impulsive  stimulated Raman excitation profiles}
\abbreviations{ISRS}

\begin{document}
	
	\begin{tocentry}
		
	\includegraphics[width=1\textwidth]{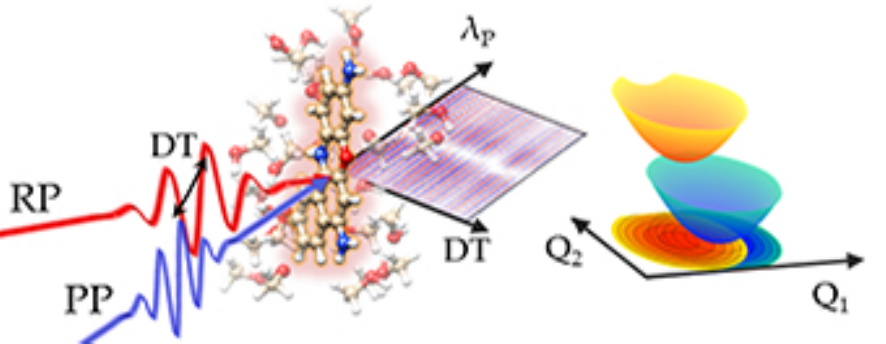}
		
	\end{tocentry}
	
	\begin{abstract}	
		Photophysical and photochemical processes are ruled by the interplay between transient vibrational and electronic degrees of freedom, which are ultimately determined by the multidimensional potential energy surfaces (PESs). Differences between ground and excited PESs are encoded in the relative intensities of resonant Raman bands, but they are experimentally challenging to access requiring measurements at multiple wavelengths under identical conditions.
		Here we perform a two-color impulsive vibrational scattering experiment to launch nuclear wavepacket motions by an impulsive pump and record their coupling with a targeted excited-state potential by resonant Raman processes with a delayed probe, generating in a single-measurement background-free vibrational spectra across the entire sample absorption. Building on the interference between the multiple pathways resonant with the excited-state manifold that generate the Raman signal, we show how to experimentally tune their relative phase by varying the probe chirp, decoding nuclear displacements along different normal modes and revealing the multidimensional PESs. Our results are validated against time-dependent density functional theory.	
	\end{abstract}

The investigation of light-induced processes is essential for the understanding of a variety of complex phenomena at the interface between physics, chemistry and biology, in which excited-state dynamics cause the transient modification of molecular properties and atomic configurations. These latter are ultimately determined by the multidimensional potential energy surfaces (PESs) describing how the excited state potential changes along the different normal modes~\cite{Zewail2000}. 
Spontaneous Raman spectroscopy plays a pivotal role for accessing the vibrational fingerprints of solid state systems or complex molecular compounds. 
Tuning the excitation of the Raman pulse in resonance with an electronic transition enables selective enhancement of the Raman cross section \cite{Long2002}. 
Since the resonance Raman (RR) cross section depends on the displacement between ground and excited potential energy surfaces along the normal coordinates, the relative intensities of the measured RR bands encode information on the potential energy surfaces. Critically, in order to extract such molecular information, several spectra have to be recorded scanning the Raman pump wavelength across the absorption profile, performing a sequence of measurements at multiple wavelengths under identical conditions, and 
with the detection of the experimental signals that is typically hampered by the overwhelming fluorescent background. 
Most importantly, spontaneous Raman can only provide differential information on excited state geometries relative
to the equilibrium configuration on the ground state.
\\
Here we introduce a method to determine the nuclear displacements between different PESs, based on an impulsive stimulated Raman scattering (ISRS) experiment, circumventing the limitations of spontaneous Raman spectroscopy.
%
ISRS exploits the joint action of two ultrashort laser fields to measure vibrational excitations in the time domain~\cite{Polli2010,Kuramochi2016,cit::IVS::kukura}. 
A femtosecond Raman pump (RP) coherently stimulates nuclear wavepackets of Raman active modes~\cite{cit::Mukamel,Mukamel_Potma_CRS}, which modulate the transmissivity of the sample and are detected by monitoring the transmission of a temporally delayed probe pulse (PP) as a function of the time delay $\Delta T$ between the RP-PP pair: Fourier transforming over $\Delta T$ enables to retrieve the Raman information in the frequency domain~\cite{cit::Champion,cit::rhuman::bacteriorhodopsin}.
Since the ISRS spectra are recorded for temporally separated Raman and probe fields, the experimental signals are not affected by nonlinear background processes generated during the overlap of the ultrashort pulses, which in contrast can plague frequency domain coherent Raman techniques \cite{cit::Agrawal,Mathies_review,Batignani2015,Virga2019,Batignani2019_adp}. In addition, thanks to the heterodyne detection, the vibrational information is engraved on the PP, hence the fluorescence and the incoherent background signals are efficiently suppressed. Importantly, when the Raman pulse is longer than the period of a normal mode, it cannot efficiently stimulate vibrational coherences, making ISRS less effective in probing high energetic modes.
In order to isolate the vibrational fingerprints on the initially populated electronic level, we employ a two-color ISRS experimental configuration, with an off-resonant Raman pulse $E_R$, and a broadband probe $E_P$, which covers the molecular response across its absorption profile, enabling to study the couplings with a targeted electronically excited level, 
by monitoring the frequency-dispersed ISRS signal as a function of the probe wavelength $\lambda_P$. 
We demonstrate how to assign the complex dependence (as a function of $\lambda_P$) of the measured mode-dependent Raman excitation profiles  to the corresponding nuclear displacements. The relative intensities of the ISRS modes are studied including chirp effects, which can critically modulate the amplitude of the time domain Raman response even in the fully off-resonance regime~\cite{Monacelli2017,Gdor2017}.
The addition of an actinic pump, temporally longer than the period of the normal modes under investigation, to the scheme proposed here allows to photoexcite the sample on a targeted electronic level~\cite{Musser2015, Fujisawa2016, Kim2020, Moran2016_perspective, Kuramochi_2019, Fumero2020} without promoting vibrational coherences. Hence, thanks to the  femtosecond resolution probing of the RP-PP pair~\cite{Takeuchi2008,Sun2014,Johnson2015,Schnedermann2016,Batignani2018}, probing the ISRS response across the excited state absorption activated upon the photo-excitation enables the extension to the mapping of the relative orientation/displacements between arbitrary excited transient PESs. Within such a scheme the RP may be tuned in resonance with the stimulated emission to enhance the ISRS cross section.
The measurement of excited state orientations by ISRS offers the chance to map complex PESs and to identify the vibrational degrees of freedom responsible for the ultrafast relaxation of the system along excited-state potentials, as in presence of dynamic Stokes shift~\cite{Brennan2017}, where measuring the time-dependent REPs would offer the chance to follow the photoexcited chromophore PES relaxation along the involved vibrational degrees of freedom, or photo-induced charge transfer events~\cite{Wang2019}.

\begin{figure}[hbtp]
	\centering
	\includegraphics[width=18cm]{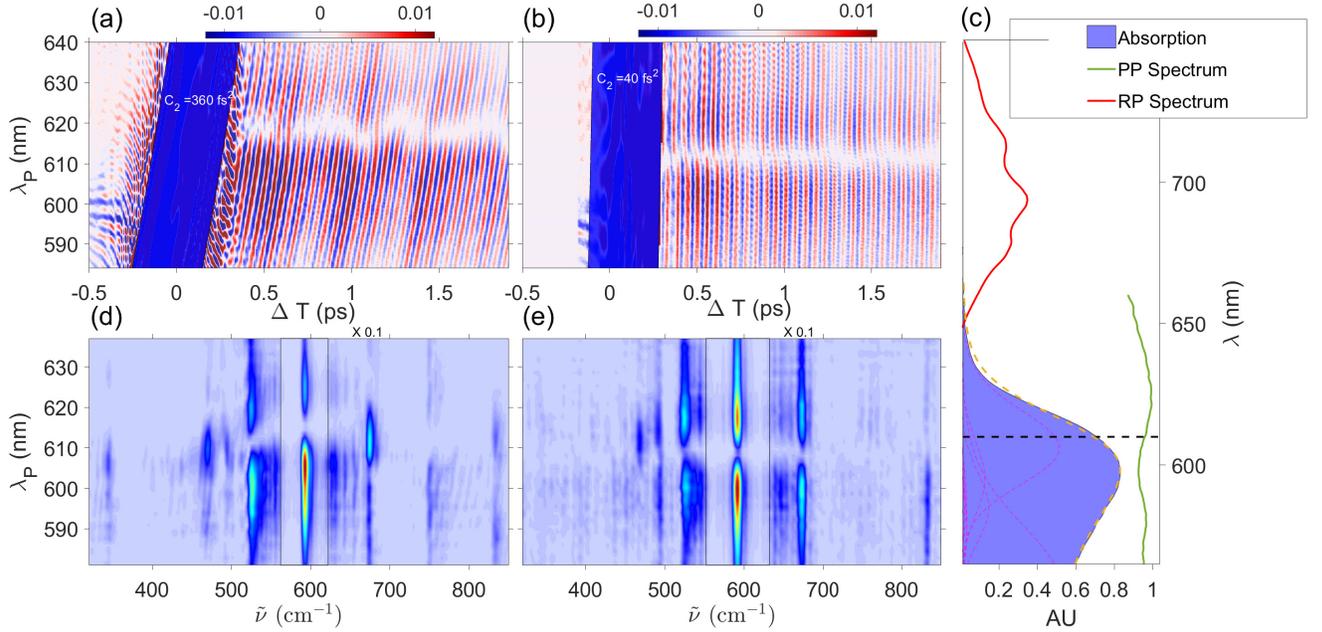}
	\caption{Broadband two-color ISRS spectra of Cresyl Violet recorded in the time domain are reported in the top panels (a-b) as a function of the probe wavelength $\lambda_P$ and the RP-PP relative delay $\Delta T$, acquired for two different values of the probe chirp (360 and 40 fs$^2$). The blue box marks the region covered by the coherent artifact. The corresponding spectra in the frequency domain (panels d-e ) can be obtained truncating the coherent artifact and Fourier transforming over $\Delta T$ and are reported in the bottom panels as a function of the wavenumber $\tilde{\nu}$ (the central region around 592 cm$^{-1}$ has been scaled by a factor 0.1 to enhance the visibility of the weaker Raman modes). 
	The absorption spectrum of the methanol dissolved sample is shown in the right panel (c) with the RP and PP spectral profiles and is compared with the one (yellow dashed line) simulated from the molecular parameters reported in table \ref{table}. A low (300 cm$^{-1}$) and a high (1400 cm$^{-1}$) frequency contributions have been included in the simulation, taking into account for low frequency contributions with vanishing polarizability derivative not stimulated by the non-resonant pump and the high frequency modes out of reach for the ISRS technique. The contributions originated by the vibronic progression are reported as magenta lines.
	The black dashed line indicates the transition to the vibrationally ground level of the excited electronic state. \label{Fig1}}
\end{figure}

The two-color ISRS response is measured for Cresyl Violet (CV) \cite{Vogel2000,Leng2003,Brazard2015,Rafiq2016,Batignani2020} dissolved in methanol, an highly fluorescent oxazine dye with a long-lived excited electronic state. 
Due to the strong fluorescence background, with a small ($\approx$ 500 cm$^{-1}$) Stokes shift, the CV spontaneous resonant Raman response cannot be explored around the system absorption profile and can be only obtained at excitation wavelengths far to the blue side of the absorption maximum \cite{Leng2003}. 
Hence CV represents an ideal candidate for testing the capabilities of our two-color ISRS setup.
Our experimental results are reported in Figure \ref{Fig1}: the temporal traces recorded in the time-domain are shown in the top panels (a-b) for two values of the probe chirp $C_2$ (360 fs$^2$ and 40 fs$^2$), while the corresponding spectra in the frequency domain, obtained upon truncating the coherent artifact (highlighted by the blue boxes) and fast Fourier transforming (FFT), are reported in the bottom panel (d-e). Since the ISRS signal is dominated by the mode centered at 592 cm$^{-1}$, the spectral region around the 592 cm$^{-1}$ peak has been scaled by a factor 0.1 in order to enhance the visibility of the other weaker Raman bands (at 345, 470, 493, 526 675, 752 and 833 cm$^{-1}$). It is worth to stress that in view of to the non-resonant nature of the Raman pump employed in the present scheme, only the Raman modes with a non-vanishing polarizability derivative can be efficiently stimulated.
As expected, the amplitude of the ISRS oscillations is enhanced around the sample absorption profile, which is reported in Figure \ref{Fig1}c. 
The intensities of the different Raman peaks show complex profiles, which vary as a function of the PP wavelength and depend on both PP chirp as well as on the specific Raman mode under consideration.
For example, the 592 cm$^{-1}$ mode shows a maximum intensity slightly red-shifted with respect to the absorption maximum for $C_2$ = 360 fs$^2$ and turns to a broad bi-lobed profile for $C_2$ = 40 fs$^2$. In contrast, the 525 cm$^{-1}$ mode shows similar bi-lobed profiles for both $C_2$ = 40 fs$^2$ and $C_2$ = 360 fs$^2$. The traces in the time-domain reveal a $\pi$ phase shift between the oscillations red and blue shifted with respect to the node with vanishing signal.

\begin{figure}[hbtp]
	\centering
	\includegraphics[width=16.8cm]{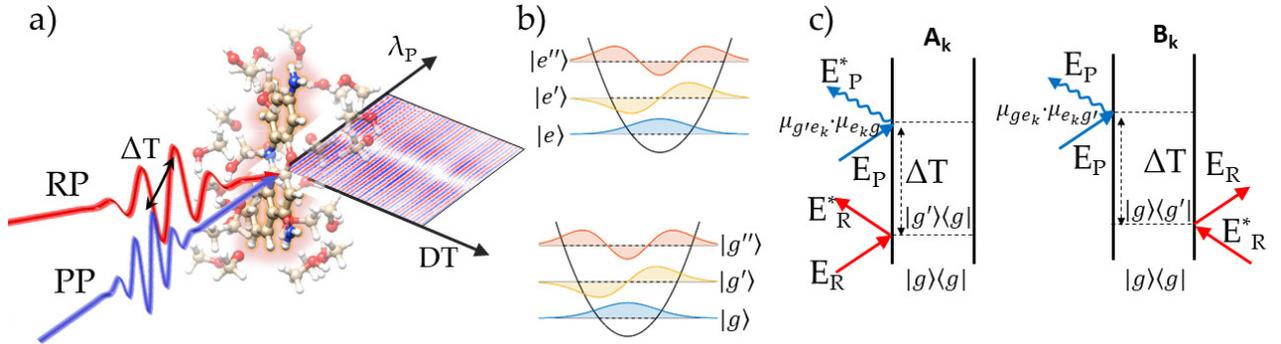}
	\caption{A sketch of the two-color ISRS experimental setup used in this work is shown in panel (a). The electronic and vibrational levels involved in the generation of the measured signal are reported in panel (b): the vibrational progression on the ground electronic state is indicated as $\ket{g}, \ket{g'}, \ket{g''}$, while the corresponding manifold in the electronic excited state is referred as $\ket{e}, \ket{e'}, \ket{e''}$. The Feynman diagrams describing the ISRS process are reported in panel (c). \label{Fig2}}
\end{figure}

\begin{figure}[hbtp]
	\centering
	\includegraphics[width=16.3cm]{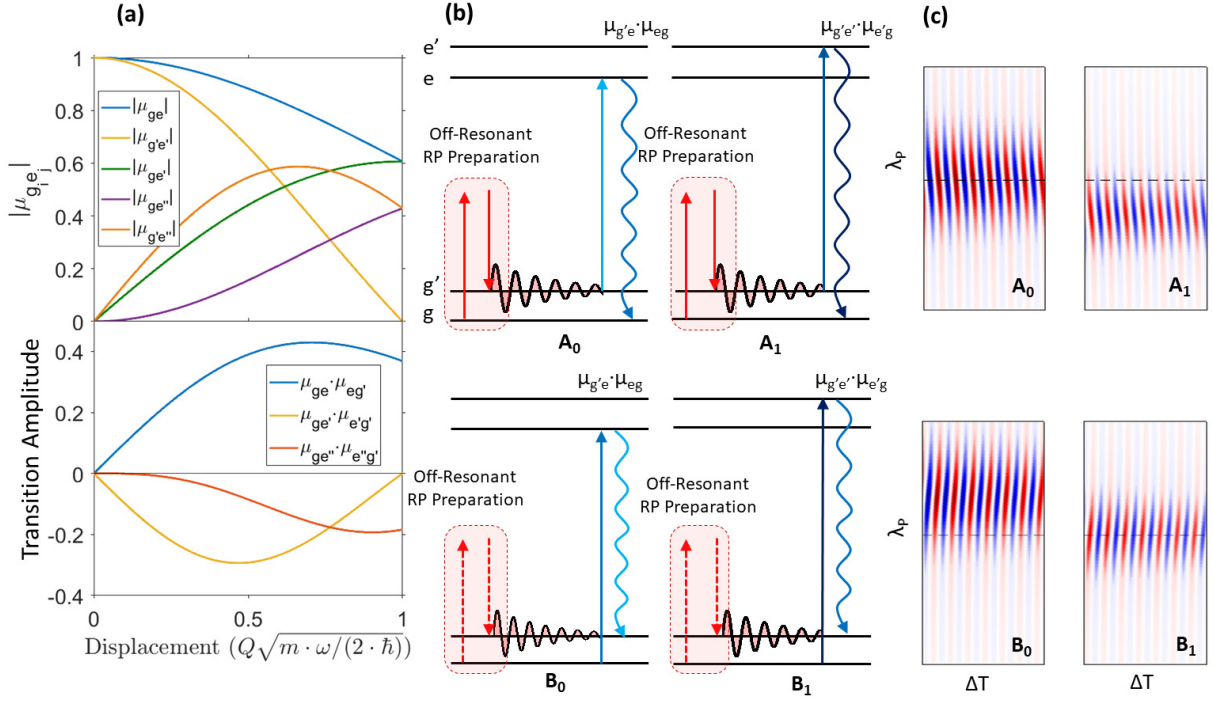}
	\caption{(a) Transition dipole moments are reported as a function of the dimensionless displacement along the normal modes. In the top panel the absolute value of single transition dipole moments are shown (the vibrational progression on the ground electronic state is indicated as $\ket{g}, \ket{g'}$, the corresponding manifold in the electronic excited state is referred as $\ket{e}, \ket{e'}, \ket{e''}$), while in the bottom panel we report the intensity factors associated to the various ISRS pathways shown in panel (b). Each factor corresponds to the product of two dipole moments. The corresponding signals in the time domain are reported in panel (c), with the horizontal dashed line that indicates the PP wavelength matching the $e-g$ transition. \label{Fig3}}
\end{figure}

To interpret this complex behaviour and extract structural information from the measurements of mode-dependent intensity profiles in CV, we studied the role of the resonant PP chirp and detection wavelength in the generation of the nonlinear signal. The third-order nonlinear polarization associated to the ISRS process can be evaluated through a perturbative expansion of the molecular density matrix in powers of the electric fields 
$E_{R/P}=E^0_{R/P}(t) e^{-i\omega_{R/P}t} + c.c.$, and the different pathways that contribute to the total response can be identified taking advantage of a diagrammatic approach \cite{cit::Mukamel,Mukamel_Rahav,Dorfman2013,Batignani2015pccp,Fumero2015}. 
Considering the pulse scheme and the energy levels reported in Figure \ref{Fig2}a-b, 
with vibrational manifolds in the electronic ground and resonant excited state indicated as $\ket{g}, \ket{g'}, \cdots$ and $\ket{e}, \ket{e'}, \cdots$, respectively, the Feynman diagrams that take into account the system response are shown in Figure \ref{Fig2}c. Since the RP is tuned off resonant with respect to the sample absorption, it selectively generates vibrational coherences only in the initially populated electronic level, the ground state in the present case ($\ket{g'}\bra{g}$ and $\ket{g}\bra{g'}$ states, corresponding to diagrams $A_k$ and $B_k$, respectively). Then the system evolves unperturbed until an interaction with the PP and a free induction decay (that leaves the system in a population state) enabling to probe the vibrational coherences after the tunable time delay $\Delta T$.
Importantly,
while in the off-resonant PP regime the state $e$ is a virtual level and $A_k$-$B_k$ destructively interfere resulting in small or vanishing signals~\cite{Monacelli2017}, in the resonant case the ket side of the density matrix is promoted to the excited state and, taking advantage of the broadband nature of the PP, it can be projected to the entire vibrational manifold, with a distribution ruled by the transition dipole moments $\mu_{nm}=\braket{\psi_{n}|\hat{\mu}|\psi_{m}}$. 
Assuming the Condon approximation~\cite{Fumero2020,Condon1947}, the intensities of the transition dipole moments 
are determined by the Franck-Condon (FC) factors and can be evaluated by computing the FC overlap integrals between the initial and final vibrational wave functions $\mu_{g_je_k}=\braket{g_j|e_k}$, which depend on the displacement of the PES along the considered normal mode. 
We adopt the harmonic approximation for the molecular potentials, according to which the vibrational wave functions are expressed by products of orthogonal shifted harmonic oscillators.
In the left panel of Figure \ref{Fig3}, we report some values of the transition dipole moments as a function of the dimensionless displacement factor $d=Q\sqrt{\frac{m\omega_0}{2\hbar}}$. 


The third-order polarizations, $P^{(3)}_{A_{g'}} (\omega,\Delta T)$ and $P^{(3)}_{B_{g'}} (\omega,\Delta T)$, taking into account the signal induced by a given vibrational mode $g'$, can be directly computed  from the diagrams in Figure \ref{Fig2} and are~\cite{Batignani2019_CISRS,Batignani2016SciRep}
\begin{align}\label{Eq:SA}
P^{(3)}_{A_{g'}} (\omega,\Delta T) = 
\sum_{e_k} \frac{\mu_{g'e_k}\mu_{e_kg}}{\hbar^3}
\int_{-\infty}^{+\infty}\int_{-\infty}^{+\infty}
d\Delta
d\omega_D
\frac{I_{RA} (\Delta,\Delta T)E_P(\omega-\Delta)G(\omega_D)}{(\Delta-\tilde{\omega}_{g'g})(\omega-\tilde{\omega}_{e_kg}-\omega_D)}
\end{align}
and
\begin{align}\label{Eq:SB}
P^{(3)}_{B_{g'}} (\omega,\Delta T) = &
\sum_{e_k} \frac{\mu_{ge_k}\mu_{e_kg'}}{\hbar^3}
\int_{-\infty}^{+\infty}\int_{-\infty}^{+\infty}
d\Delta
d\omega_D
\frac{I_{RB} (\Delta,\Delta T)E_P(\omega+\Delta)G(\omega_D)}{(-\Delta-\tilde{\omega}_{gg'})(\omega-\tilde{\omega}_{e_kg'}-\omega_D)}
\end{align} 
$\tilde{\omega}_{ij}=\omega_{ij} -i\gamma_{ij}$, $\omega_{ij}=\omega_{i}-\omega_j$ indicates the energy difference between i-j levels, 
$\Delta$ is an integration variable (with the maximum contribution centered at $\tilde{\omega}_{g'g}$), 
$G(\omega_D)$ is the inhomogeneous broadening Gaussian function and	$\gamma_{ij}=\tau_{ij}^{-1}$ is the  dephasing rate of the $\ket{i}\bra{j}$ coherence.
The $I_{RA} (\Delta,\Delta T)$ and $I_{RB} (\Delta,\Delta T)$ terms in Eqs. \ref{Eq:SA}-\ref{Eq:SB} represent the preparation function of the  $\ket{g'}\bra{g}$ and $\ket{g}\bra{g'}$ vibrational coherences generated by the non-resonant Raman pulse and, 
under off-resonant RP excitation, they are purely real functions. 
Importantly, Raman modes with a period much shorter than the pump duration will exhibit an inefficient stimulation of the vibrational coherences, resulting in small amplitudes of the ISRS peaks \cite{cit::IVS::kukura}.
The summation over $e_k$ in Eqs. \ref{Eq:SA}-\ref{Eq:SB} takes into account for the PES 1-dimensional projection on the $g'$ vibrational subspace, and, in order to retrieve the desired total nonlinear polarization of the system, Eqs. \ref{Eq:SA}-\ref{Eq:SB} should be summed over all the $g'$ ground state normal modes under consideration, accordingly to
$$
P^{(3)}_{A/B}(\omega,\Delta T) =\sum_{g'} P^{(3)}_{A_{g'}/B_{g'}} (\omega,\Delta T)
$$
Details on the derivation of Equations \ref{Eq:SA}-\ref{Eq:SB} and expressions for the $I_{RA} (\Delta,\Delta T)$ and $I_{RB} (\Delta,\Delta T)$ functions are reported in the Supporting Information.
For simplicity, the calculation can be performed assuming that the central frequency $\omega_P$ of the chirped probe pulse arrives at $t=0$ and that the RP is centered at $t=-\Delta T$ (i.e. shifted at negative time delays and preceding the PP). Under such assumptions the RP-PP fields of Eqs. \ref{Eq:SA}-\ref{Eq:SB} can be expressed in the frequency domain as
\begin{equation}\label{Eq: Fields}
E_R(\omega)=E_R^{(0)}(\omega)e^{-i\omega_R \Delta T}
\hbox{, }
E_P(\omega)=E_P^{(0)}(\omega)e^{i\sum_n C_n (\omega-\omega_P)^n}
\end{equation}
where $E_{R/P}^{(0)}(\omega)$ are positive real functions representing the square root of the RP-PP spectra,
$C_2$ is the group delay dispersion taking into account the linear chirp of the PP and
$C_n$ indicate the $n^{th}$ higher order dispersion terms.
\\
When the PP do not vary across the sample, the third-order nonlinear polarization, which depends on the PP temporal and spectral profile, is constant and hence the ISRS responses $S(\omega,\Delta T)$-$S(\omega,\Omega)$, in the time and frequency domains respectively, can be calculated as:
$$
S(\omega,\Delta T) \propto - \Im \left[
\left(
P_A^{(3)}(\omega)+P_B^{(3)}(\omega)\right)
E_P^{(0)*}(\omega)
\right]
$$
$$
S(\omega,\Omega) = 
\int_{-\infty}^{+\infty}d\Delta T e^{+i\Omega \Delta T}
S(\omega,\Delta T)
$$
where $\Im (f)$ indicates the imaginary part of $f$.\\
However, we note that, due to the absorption of the $E_P$ field during its propagation,
the polarizations $P_A^{(3)}(\omega)$ and $P_B^{(3)}(\omega)$ are not constant and decay along the beam path within the sample.  
For this reason, a quantitative evaluation of the ISRS signal requires to numerically integrate the PP spectral profile over the sample length, by using the coupled wave equations~\cite{cit::Agrawal,Batignani2019}
\begin{equation}\label{Eq:Propagation}
\left\{
\begin{array}{ll}
\frac{\partial E_P(\omega,z)}{\partial z}=  \frac{i\omega}{c} \left[P_A^{(3)}(z,\omega)+P_B^{(3)}(z,\omega)\right]-\frac{\alpha(\omega)}{2} E_P(\omega,z)\\
P_A^{(3)}(z,\omega)=
\sum_{e_k} \frac{\mu_{g'e_k}\mu_{e_kg}}{\hbar^3}
\int_{-\infty}^{+\infty}\int_{-\infty}^{+\infty}d\Delta d\omega_D
\frac{I_{RA} (\Delta)E_P(\omega-\Delta,z)G(\omega_D)}{(\Delta-\tilde{\omega}_{g'g})(\omega-\tilde{\omega}_{e_kg}-\omega_D)}
\\
P_B^{(3)}(z,\omega)=
\sum_{e_k} \frac{\mu_{ge_k}\mu_{e_kg'}}{\hbar^3}
\int_{-\infty}^{+\infty}\int_{-\infty}^{+\infty}d\Delta d\omega_D
\frac{I_{RB} (\Delta)E_P(\omega+\Delta,z)G(\omega_D)}{(-\Delta-\tilde{\omega}_{gg'})(\omega-\tilde{\omega}_{e_kg'}-\omega_D)}
\end{array}
\right.
\end{equation}
where $\alpha(\omega)$ indicates the frequency dependent attenuation coefficient.
The ISRS signal $S(\omega,\Delta T)$ in the time domain can be finally evaluated as the normalized difference between the transmitted PP spectrum in presence  and in absence of the RP induced nonlinear polarization.

To interpret the experimental results it is worth to consider separately the different pathways, contributing to the $e_k$ summations (Eqs. \ref{Eq:SA}-\ref{Eq:Propagation}), that generate the polarizations $P_A$ and $P_B$. 
From a diagrammatic perspective, this is equivalent to recast the Feynman diagrams shown in Figure \ref{Fig2}c to the corresponding ones in the energy level scheme, which are reported in Figure \ref{Fig3}b (we have explicitly shown only the $e$ and $e'$ vibrational levels in the electronic excited state). Diagrams in Figure \ref{Fig3}b help to directly visualize the spectral components of the Raman and probe fields contributing to the experimental signal generation at a given PP wavelength.
Importantly, at odd with fully non-resonant ISRS processes, where the total response of the system is generated by the interference between pathways involving interactions with different probe pulse spectral components (red-shifted and blue shifted with respect to the probed wavelength)~\cite{Zhou1999,Gdor2017,Batignani2019_CISRS}, in the resonant case the signal at a given probe wavelength is generated by the interference between processes which share interactions with the same probe color, but with a different state in the excited state manifold, hence encoding information on the nuclear displacements between different potential energy surfaces.
In particular, the  $A_0$/$A_1$ processes are generated from an interaction with a PP component red-shifted by one vibrational quantum with respect to the probed wavelength, and originate from the transitions $e\rightarrow g$, $e'\rightarrow g$, hence resulting in
contributions centered at $\omega_{eg}$ and $\omega_{e'g}$, respectively.
On the contrary $B_0$ and $B_1$ generate contributions  centered around $\omega_{eg'}$ and $\omega_{e'g'}$ and originate from interactions with spectral components of the PP blue-shifted with respect to the probed wavelengths. 
Interestingly, while $A_0$ and $B_0$ generate an ISRS response centered at different PP wavelengths, they originate from quantum pathways involving the same excited state $e$, resulting in ISRS signals oscillating in phase with each other (Figure \ref{Fig3}c).
In contrast, both $A_0$ and $B_1$ generate an ISRS response centered at the same transition frequency ($\omega_{eg}$). However, the amplitude of the corresponding oscillations is ruled by different dipole moments, namely $\mu_{g'e}\mu_{eg}$ for $A_0$ and  $\mu_{g'e'}\mu_{e'g}$ for $B_1$. Critically, as shown in Figure \ref{Fig3}a, $\mu_{g'e}\mu_{eg}$ and $\mu_{g'e'}\mu_{e'g}$ have opposite signs and hence they result in a destructive interference between the $A_0$ and $B_1$ processes. This elucidates the vanishing ISRS amplitude observed at the transition frequency $\omega_{eg}$ in Figure \ref{Fig1}b,e.
Similarly, the relative sign between the $A_1$ and $B_0$ processes, red and blue shifted by one vibrational quantum with respect to the $eg$ transition, rationalizes the $\pi$ phase shift observed in the time domain traces around the node (Fig. \ref{Fig1}a-b).
It is worth to stress that the interference between $A_0$ and $B_1$ pathways can strongly affect the detection of the Raman modes with small displacements between ground and excited PESs, i.e. the weakest Raman modes. In fact, as shown in the bottom panel of Figure \ref{Fig3}a,  
for a small displacement ($d\lesssim 0.25$), the  $\mu_{g'e}\mu_{eg}$ and $\mu_{g'e'}\mu_{e'g}$ transition amplitudes are close in absolute value to each other.
A convenient way to avoid the destructive interference between different quantum pathways can be derived from the observation that $A_0$ and $B_1$ are generated by interactions with different PP spectral components (red and blue separated by one vibrational quantum, with respect to the probed wavelength), which therefore can be shifted in time introducing a chirp in the PP. 
For a linearly chirped PP, the arrival time of a given frequency $\omega$ varies as $t=2 C_2 \left(\omega-\omega_P\right)$. The effective detection times of the vibrational coherences
are equal to the average between the arrival times of the two PP frequencies involved in the nonlinear process, and hence they correspond to 
$t_{eff}^{A_k}(\omega)=2 C_2 \left(\omega-\omega_P-\frac{\omega_{g'g}}{2}\right)$, $t_{eff}^{B_k}(\omega)=2 C_2 \left(\omega-\omega_P+\frac{\omega_{g'g}}{2}\right)$,  respectively.
This delay introduces a relative phase between the $A_k$ and $B_k$ processes, equal to $\Delta\phi_{B-A}=\omega_{g'g'}\left(t_{eff}^{B_k}(\omega)-t_{eff}^{A_j}(\omega)\right)=2 C_2\omega_{g'g}^2$, which can be experimentally tuned by means of the probe chirp.  


The above analysis clarifies that a wealth of information on the excited state potentials is encoded in the dependence of the ISRS signal on the probe wavelength. This dependence is complicated by multiple interfering optical processes but can be controlled by varying the chirp of the PP.
Following this strategy, we determined the nuclear displacements between the ground and excited PESs from the measured, mode-dependent Raman excitation profiles (REPs), i.e. the change in the intensities of the individual ISRS bands as a function of the probe wavelength.
\begin{figure}[hbtp]
	\centering
	\includegraphics[width=18cm]{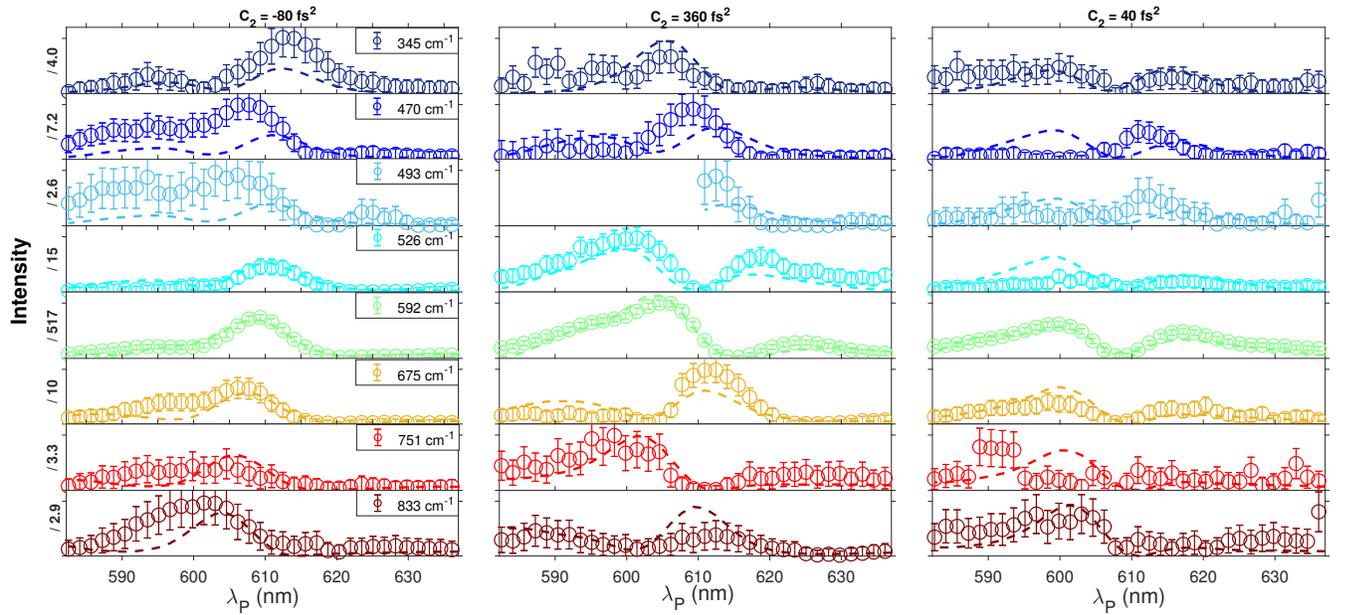}
	\caption{Normalized Cresyl Violet ISRS Raman excitation profiles measured for different vibrational modes. The REP intensities have been obtained by fitting the frequency domain traces (Figures \ref{Fig1}d-e) with a sum of Lorentzian profiles, evaluating the areas of the different Raman bands. 
	The data (circles) and the model (solid lines) are compared as a function of the $\lambda_P$ probe pulse wavelength for three different values of the probe chirp (reported on top of each column). All the y-axes range from 0 to 1.2 and the absolute intensities of each mode have been scaled by the factors reported on the axis. \label{Fig4}}
\end{figure}
In Figure \ref{Fig4}, we report the cresyl violet REP $S(\lambda_P,\tilde{\nu_j})$ for different Raman modes ($\tilde{\nu_j} = $ 345, 470, 493 ,525, 592, 675, 751 and 832 cm$^{-1}$).  $S(\lambda_P,\tilde{\nu_j})$ have been extracted by fitting the ISRS maps in the frequency domain (Figure \ref{Fig1}d-e) as the sum of Lorentzian profiles, which have been used to preliminary identify the normal mode frequencies. 
The REPs have been evaluated for three different values of the probe chirp (-80, 40 and 360 fs$^2$) to tune the relative phase between $A_0$-$B_1$ processes: the experimental results are in good agreement with signals modeled using Eqs. \ref{Eq:SA}-\ref{Eq:Propagation},  which have been numerically integrated using small steps of $dz = $ 12.5 $\mu m$,  considering the sample absorption and the nonlinear contributions separately~\cite{cit::Agrawal}.
Since the $S(\lambda_P,\tilde{\nu_j})$ signal is evaluated over  a broad spectral range,  the PP chirp has been included in the simulation considering also  the third order dispersion coefficient $C_3$ (details on the calibration are reported in Methods and in the Supporting Information).
In order to retrieve the displacements of the excited PES along the normal modes,
 Eqs. \ref{Eq:SA}-\ref{Eq:Propagation}
have been used to fit the experimental $S(\lambda_P,\tilde{\nu_0})$ traces globally over the datasets measured at different PP chirps, considering fixed normal mode frequencies and dephasing times. 
The obtained normal mode frequencies $\tilde{\nu}_{g'g}$ and displacements $d=Q\sqrt{\frac{m\omega_0}{2\hbar}}$ are summarized in table \ref{table}, while the projection of ground and excited state potentials along two normal modes (592 and 675 cm$^{-1}$) are reported in Figure \ref{Fig5}.  Importantly, our results indicate a high displacement ($d=0.64\pm0.04$) of the excited PES along the 592 cm$^{-1}$ coordinate, pointing to an elongation of the Oxazine ring in the excited state (Fig. \ref{Fig5}), due to a reorganization of the electronic density in the excited state (as confirmed by the electronic density difference map reported in the Supporting Information as Fig. S5), and small $d$ ($\leq 0.25$) along the other investigated vibrational degrees of freedom.
We note that the values reported in table \ref{table} are lower than the displacements obtained by spontaneous Raman spectroscopy for CV dissolved in water~\cite{Leng2003}, as expected in view of the asymmetrically blue shifted absorption spectrum  of the aqueous CV solutions (reported in the Supporting Information). 
In Fig. \ref{Fig1}c, we show a comparison between the measured sample absorption spectrum  and the one calculated using the ISRS retrieved displacement, where  two additional contributions (at 300 and 1400 cm$^{-1}$), that take into account for the Raman mode not accessed by the present ISRS experiment, have been included in the simulation to fit the experimental data. Interestingly, such a comparison indicates that CV has high-frequency displaced normal modes outside the frequency range investigated in the present work, in agreement with the results reported for CV dissolved in aqueous solution ~\cite{Leng2003}. 
\begin{table}
	\begin{tabular}{|c|c|c|c|}
		\hline
		$\tilde{\nu}_{g'g}$ (cm$^{-1}$) & $d=Q\sqrt{\frac{m\omega_0}{2\hbar}}$ (DFT) & $d=Q\sqrt{\frac{m\omega_0}{2\hbar}}$ (experimental) & $Q (a.u.)$ \\
		\hline
		\hline
		345 (2) & 0.27 & 0.18 (0.03) & 0.063\\
		\hline
		470 (2) & 0.24 & 0.21 (0.04) & 0.059\\
		\hline
		493 (2) & 0.13 & 0.15 (0.03) & 0.040\\
		\hline
		526 (2) & 0.18 &  0.26 (0.03) & 0.071\\
		\hline
		592 (2) & 0.60 & 0.63 (0.04) & 0.13\\
		\hline
		675 (2) & 0.19  & 0.19 (0.03) & 0.044\\
		\hline
		752 (2) & 0.11  & 0.17 (0.03) & 0.039 \\
		\hline
		833 (2) & 0.17  & 0.17 (0.03) & 0.041 \\
		\hline
	\end{tabular}
	\caption{Peak positions and displacements between ground and excited state PESs extracted by fitting the ISRS Raman excitation profiles are reported with the corresponding 90\% confidence intervals. The results are compared with the displacements obtained by TD-DFT calculations performed with CAM-B3LYP functional and the 6-311++G(2d,2p) basis set. In the last column the extracted detected displacements are reported in atomic units.}\label{table}
\end{table} 
\\
The experimental results are compared with ab initio time dependent density functional theory (TD-DFT) calculations performed with CAM-B3LYP functional~\cite{Yanai2004} and the 6-311++G(2d,2p) basis set. Interestingly, the TD-DFT calculations reported in table \ref{table} and the ISRS extracted disaplcements are in good agreement, confirming the capability of the presented approach to access the details of the excited state PES, also for the normal modes with small displacements.
A detailed comparison between the ISRS experimental results and the TD-DFT calculations is reported in the Supporting Information.
\begin{figure}[hbtp]
	\centering
	\includegraphics[width=16cm]{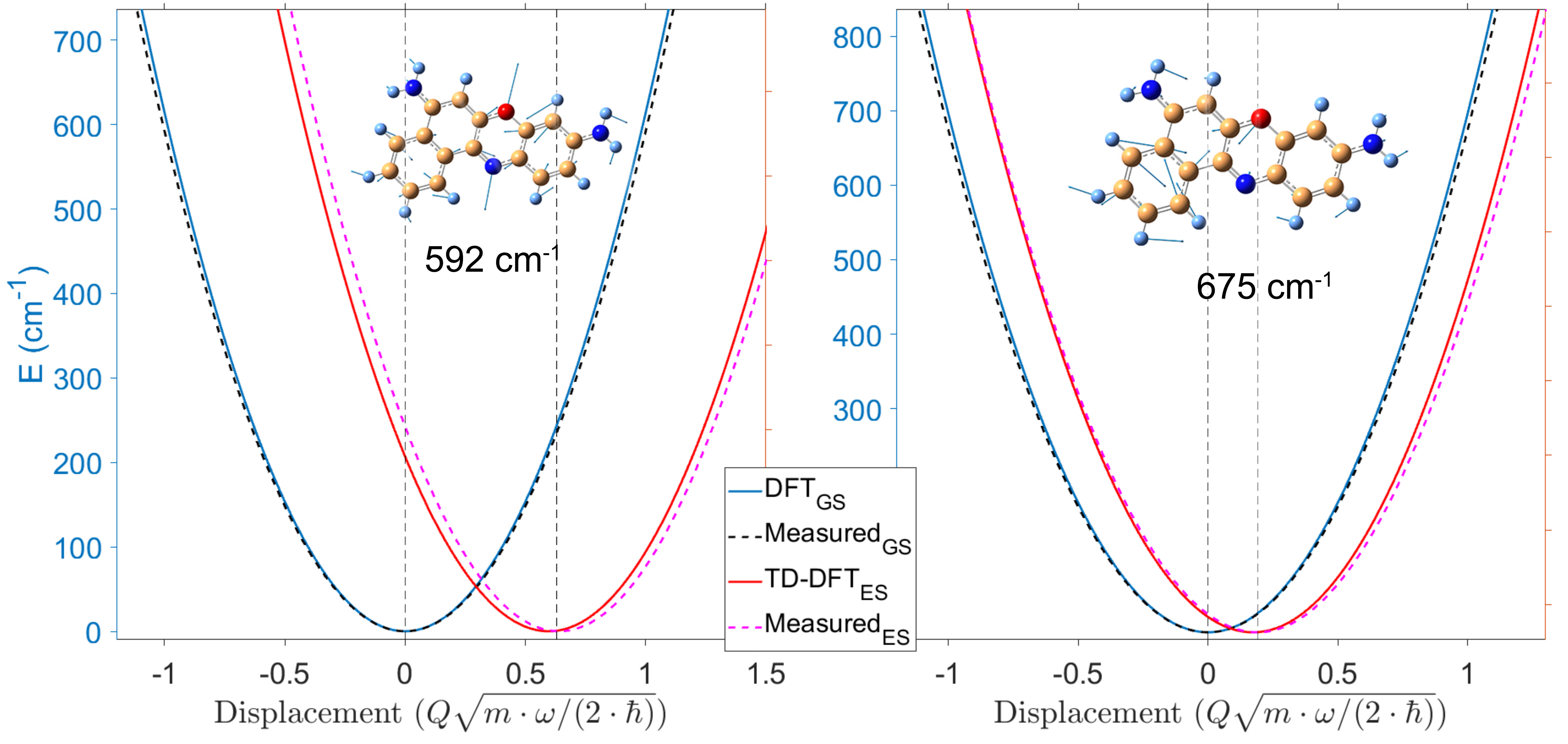}
	\caption{Cresyl Violet potential energy surfaces along two normal modes (592 and 675 cm$^{-1}$) measured by mapping the ISRS Raman excitation profiles. The experimental results are compared with time dependent density functional theory calculations performed with CAM-B3LYP functional and the 6-311++G(2d,2p) basis set. The vertical dashed lines indicate the measured minima of ground and excited state parabolas. The excited state potentials are vertically offset by a constant factor. \label{Fig5}}
\end{figure}

In conclusion, we have investigated the ISRS response in the presence of an off-resonant RP and a broadband resonant PP. 
A diagrammatic treatment of the pathways concurring to the signal generation has been employed to analyze the data.
Taking advantage of the heterodyne background and fluorescence-free detection generated on the broadband probe pulse, Raman excitation profiles can be recorded in a single measurement,  uncovering the vibrational response of systems that cannot be accessed by spontaneous or frequency-domain approaches.
The interaction with the resonant PP enables the projection of the molecular density matrix to the entire vibrational manifold in the excited electronic state, resulting in the interference between quantum pathways that concur to the generation of the experimental signal. Utilizing this molecular description, we have shown how to retrieve detailed information on the dipole moments over the different recorded Raman modes.
As a benchmark of the proposed theoretical model and experimental scheme, we applied the two-color ISRS setup to map the  PESs of cresyl violet dissolved in methanol, determining the nuclear displacements of excited state PESs along the different monitored normal modes.
Furthermore, our results establish a convenient experimental protocol, based on the use of a resonant chirped PP, to enhance the ISRS cross section of low scattering Raman modes.  
This protocol can be exploited for the identification of mode displacements for PESs involved in relaxation dynamics, being relevant for accessing the reaction coordinates that rule the initial stages of photoreactions. By adding an actinic pump beam indeed our approach can be extended straightforward for mapping the relative orientation between different excited transient PESs.


\section{Methods}
\subsection{Experimental setup}
The  experimental setup exploited for the measurements is based on a 1 kHz repetition rate Ti:sapphire laser source that generates 3.5 mJ, 35 fs pulses at 800 nm. 
The vertically polarized RP is synthesized by a non-collinear optical parametric amplifier (NOPA) that produces 15 fs broadband pulses centered at 680 nm and its compression is controlled by a pair of chirped mirrors~\cite{Zavelani-Rossi_2001}. 
The 10 nJ vertically polarized probe pulse is a white light continuum~\cite{cit::Agrawal} generated by focusing part of the source pulse on a sapphire plate and filtering the 800 nm component by means of a shortpass filter. 
The PP chirp can be reduced by a second pair of chirped mirror or increased by introducing glass windows of different widths along the beam path. Both the pulses are focused on a 500 $\mu$m glass cuvette containing the cresyl violet solution and then, the transmitted PP is frequency dispersed by a spectrometer onto a CCD device. A synchronized chopper at 500 Hz blocks alternating RP pulses in order to measure the transmitted PP in both the presence and absence of the RP excitation. In such a way, the ISRS signal $S(\omega,\Delta T)$ can be extracted as 
the normalized difference between the transmitted PP spectrum in presence ($I_{RP-On}(\omega,\Delta T)$) and in absence ($I_{RP-Off}(\omega)$) of the RP:
$$
S(\omega)=\frac{I_{On}(\omega,\Delta T)-I_{Off}(\omega)}{I_{Off}(\omega)}
$$
The measurement of the PP chirp and the calibration of the $C_2$ and $C_3$ terms in Eq. \ref{Eq: Fields} can be obtained monitoring both the coherent artifact generated inside the sample as well as the slope of the coherent oscillations recorded in the time domain due to the 1040 cm$^{-1}$ Raman mode of the solvent (details are reported in the Supporting Information). Taking advantage of the small period of such mode ($\approx$ 30 fs) this approach represents a convenient way to measure the probe pulse chirp simultaneously with the ISRS measurement. In addition, the chirp is directly measured in the very same spatial region where the ISRS signal is generated, hence directly taking into account for the dispersion of the probe pulse during the propagation in the sample cuvette. Further details on the chirp calibration are reported in \cite{Monacelli2017,Batignani2019_CISRS}.
The ISRS spectra of CV have been acquired for a 6.5 ps temporal window ($\Delta T$ spans from -0.5 ps to 6 ps), much higher than the dephasing time of the vibrational coherences ($<$ 3 ps for all the modes under consideration), with a 13 fs sampling interval. 
Further details on the experimental scheme are reported in~\cite{Ferrante2020}.
\subsection{Density Functional Theory calculations}
DFT and TD-DFT calculations reported in Figure \ref{Fig5} have been performed with CAM-B3LYP~\cite{Yanai2004} functional and 6-311++G(2d,2p) basis set, by using the Gaussian 09 software package. Upon optimization of the ground state geometry by DFT, the normal modes eigenvectors calculated in the ground state have been exploited to generate displaced geometries along the normal modes under consideration, which in turn have been exploited to calculate the excited state electronic potential by TD-DFT, with the same level theory. Further details are provided in the Supporting Information.

\begin{acknowledgement}
G.B. and T.S. acknowledge the  ``Progetti di Ricerca Medi 2019”grant by Sapienza Universitá di Roma and the 
 ``MESPES: Mapping Excited State Potential Energy Surcafes” project by Italian Super Computing Resource Allocation. G.B.
acknowledges Massimiliano Aschi for helpful support with computational resources and Emanuele Mai for constructive
discussions. S.M gratefully acknowledges the support of NSF through Grant CHE-1953045. This project has received
funding from the PRIN 2017 Project, Grant No. 201795SBA3-HARVEST, and from the European Union’s
Horizon 2020 research and innovation program Graphene Flagship under Grant Agreement No. 881603.	
\end{acknowledgement}

\begin{suppinfo}
	
	The Supporting Information is available free of charge and includes: 
	    (i) Cresyl violet ISRS spectra obtained with a negatively chirped probe pulse;
		(ii) a detailed derivation of the ISRS signals;
		(iii) details on the DFT and TD-DFT calculations;
		(iv) details on the calibration of the PP chirp;
		(v) absorption spectrum of CV.
	
\end{suppinfo}

\bibliography{biblio3}

\end{document}